\documentclass[a4paper,reqno]{amsart}
\usepackage{endnotes}
\usepackage[dvips]{graphicx}
\usepackage{setspace} 
\usepackage[obeyspaces]{url} 
\author{Frank G. Borg}
\thanks{Jyv{\"a}skyl{\"a} University, Chydenius Institute, POB 567, FIN-67101 Karleby, Finland. Email: \url{borgbros@netti.fi}.}
\title[An inverted pendulum ...]{An inverted pendulum with a springy control as a model of human standing}

\begin{document}

\maketitle


\begin{abstract}

The normal and the inverted pendulum continue to be one of the main physical models and metaphors in science. The inverted pendulum is also a classic study case in control theory. In this paper we consider a special demonstration version of the inverted pendulum which is controlled via a spring. If the spring constant is below a critical level the springy control will be unstable and the pendulum will be kept from falling only by exercising a dynamically varying control. This situation resembles the case of human bipedal quiet standing with the Achilles tendon serving as the spring. 

\end{abstract}

\section{Introduction}

With little exaggeration one can say that one of the most important contributions to physics ever made was by Christiaan Huygens (1629-1695) through his investigations of the pendulum.\endnote{C. Huygens, \textit{Horologium oscillatorium} (Paris, 1673). (English translation by R. J. Blackwell, Iowa State Press, 1986.) Despite that Newton was not in a habit of commending other researchers he held Huygens in highest esteem and referred to him as ''Summus Hugenius'' though Huygens would disagree with Newton's theory of gravity. A recent book pays homage to the pendulum: G. L. Baker and J. A. Blackburn, \textit{The pendulum. A case study in physics} (Oxford University Press, 2005).} Among other things they inspired Newton on his road to the Principia.  Subsequently also the inverted version of the pendulum has been used in order to demonstrate a number of fundamental topics in physics, such as instability and chaos, in many papers too numerous to be listed here. A quick online search using the key phrase ''inverted pendulum'' yielded no less than 22 papers from 1965 onwards in the \textit{American Journal of Physics} alone.\endnote{Of these we may mention Duchesne, C. W. Fischer, C. G. Gray, and K. R. Jeffrey, ''Chaos in the motion of an inverted pendulum: An undergraduate laboratory experiment,'' Am. J. Phys. 59, 987-992 (1991), and J. A. Blackburn, H. J. T. Smith, and N. Gr{\o}nbech-Jensen, ''Stability and Hopf bifurcations in an inverted pendulum,'' Am. J. Phys. 60 (10), 903-908 (1992). A large listing of ''pendulum references in physics education'' compiled by C. Gauld and M. R. Matthews can be found at the web site 
\url{http://www.arts.unsw.edu.au/pendulum/bibliography.html} hosted by the University of New South Wales.} We will describe in the present paper a controlled inverted pendulum version inspired by a biomechanical model of human standing which thus may be of additional interest to physics students. Human quiet standing is in fact one of the classical problems of biomechanics and it has occasionally given rise to lively debates about the nature of the supposed physiological control mechanism of balance.\endnote{\label{DEBATE} Of the recent contributions to the discussion on the nature of the balance control we may mention the following representative papers: D. A. Winter et al., ''Ankle muscle stiffness in the control of balance during quiet standing.'' Journal of Neurophysiology 85, 2630-2633 (2001); R. J. Peterka, ''Postural control model interpretation of stabilogram diffusion analysis,'' Biological Cybernetics 82, 335-343 (2000); P. G. Morasso and M. Schieppati, ''Can muscle stiffness alone stabilize upright standing?,'' Journal of Neurophysiology 83, 1622-1626 (1999); I. D. Loram, S. Kelly and M. Lakie, ''Human balancing of an inverted pendulum: is sway size controlled by ankle impedance?,'' Journal of Physiology 532, 879-891 (2001); P. Gatev et al., ''Feedforward ankle strategy of balance during quiet stance in adults,'' Journal of Physiology 514.3, 915-928 (1999). One of the controversial issues is whether balance is controlled by ''passive'' stiffness, or whether active feedback/forward control is necessary.} Everyone though seems to agree that, for quiet standing, the human subject can be described to a good approximation by an inverted pendulum pivoted at the ankle joints, and especially so if one restricts the attention to the forward-backward (anterior-posterior) sways. If we consider the human inverted pendulum (HIP), then during normal quiet standing the center of gravity of the person will on the average be a few centimeters ahead (anterior) of the ankle joints; that is, the person leans slightly forward. One implication is that in order to keep the person from falling only the leg muscles on the back side of the leg need to be active, pulling the body backwards against gravity. (Naturally there must be other postural muscles involved which will keep the trunk, legs and head from moving relative to each other, but they have a more static role during quiet standing. Thanks to the postural muscles ''freezing'' the degrees of freedom of the system we may use the HIP approximation.) A simplified model of the situation is shown in Fig. 1. The muscles responsible for the backward torque vis-\`a-vis the ankle joint are the so called plantar flexors making up the \textit{triceps surae} consisting of the Soleus and the lateral and medial heads of Gastrocnemius. These muscles are all joined at the Achilles tendon which in turn is attached to the heel bone (calcaneus). The interesting point here is that the muscle-tendon system cannot lock the HIP into a steady position, instead the person sways back and forth with an amplitude, in term of the center of mass, of the order of 10-20 mm. Fig. 2 gives and example of a the time series (stabilogram) of the anterior/posterior  center of pressure (A/P COP) during quiet standing, showing the erratic nature of the swayings. The stabilogram has been measured with a force plate, which is a rectangular plate with force transducers in each corner sensing the vertical forces $F_i$.\endnote{Note that the ingenious force plate described by R Cross , ''Standing, walking, running, and jumping on a force plate,'' Am. J. Phys. 67 (4), 304-309 (1999), is not quite suitable for quasistatic balance measurements which record the center of pressure. One has to employ e.g. strain-gauge force transducers for this purpose. In principle a minimal tripod system could suffice using only two transducers if the third corner rests on a ball bearing. When making quiet standing measurements the following standard test conditions are recommended: a stance with 30$^\circ$ between the medial sides of the feet and ca 2 cm heel-to-heel distance (clearance); arms relaxed at the sides; the participant is instructed to fix the eyes on a spot on the wall ca 3 m away (eyes open condition, EO). For clinical aspects of balance measurements see P.-M Gagey and B. Weber, \textit{ Posturologie. Regulation et d{\'e}r{\`e}glements de la station debout}
(Masson, 1999), 2nd ed., and \url{http://perso.club-internet.fr/pmgagey/}.} If the rear transducers are numbered 1 and 2, and the front transducers by 3 and 4, then the A/P COP coordinate $u$ is given by,

\begin{equation}
\label{eq:COP}
u = \frac{b}{2} \cdot \frac{F_3 + F_4 - F_1 - F_2}{F_3 + F_4 + F_1 + F_2}. 
\end{equation}

Here $b$ denotes the distance between the front and rear transducers and the A/P COP variable $u$ is measured from the center of the force plate. As will be shown below, the A/P COP coordinate $u$ is closely related to the torque acting vis-\`{a}-vis the ankle joints. Thus, Fig. 2 demonstrates the incessant modulating activity of the plantar flexors. In fact, stabilograms like that in Fig. 2 have suggested a comparison with Brownian motion\endnote{J. J. Collins and C. J. De Luca, ''Random walking during quiet standing,'' Phys. Rev. Lett. 73 (5), 764-767 (1994).} (of a trapped particle), and the use of methods from statistical physics, such as the Fluctuation Dissipation Theorem\endnote{M. Lauk et al., ''Human balance out of equilibrium: Nonequilibrium statistical mechanics in posture control,'' Phys. Rev. Lett. 80, 2, 413-416 (1998); C. C. Chow and J. J. Collins, ''Pinned polymer model of posture control,'' Phys. Rev. E 52 (1), 907-912 (1995).}, and it has even lead to an application of the notion of Stochastic Resonance.\endnote{A. Priplata et al., ''Noise-enhanced human balance control,'' Phys. Rev. Lett. 89 (23), 238101 (2002).} Although these contributions from physics have brought new methods of analysis into posturology, such as the Stabilogram Diffusion Analysis (based on the Detrended Fluctuation Methods),\endnote{For a review of some of these methods see F. G. Borg, ''Review of nonlinear methods and modelling,'' physics/0503026 (2005); ''Random walk and balancing,'' physics/0411138 (2004).} the reason for the apparent chaotic swayings have not been much touched upon. An important factor that has often been overlooked is the compliance of the Achilles tendon, as has been emphasized in the recent discussion.$^{\ref{DEBATE}}$ With a stiff tendon the muscles could in principle lock the person in a forward leaning position and the observed swaying would be attributed to fatigue, or some sort of tremor; however, with an enough compliant tendon such an equilibrium position is unstable. In order to demonstrate the phenomenon one can use an inverted pendulum as described in section {\ref{DEVICE}}. The pendulum is manually operated via the wire which runs by a pulley (whose placement corresponds roughly to the heel bone) and connects with the pendulum by a spring. The task is to try to keep the pendulum in a slightly forward leaning position by pulling from the wire. With a spring whose spring constant is below a critical value the forward leaning position is unstable. A small disturbance forward will lead to gravity taking over and the pendulum falling forward; a small disturbance backwards again will lead to the spring taking over and a toppling backwards. This situation forces one to employ an oscillatory mode of control in order to keep the pendulum from toppling. For the demonstration device it is easy to measure the parameters, such as the spring constant, but in the physiological case it is more involved. Yet measurements in vivo, using ultrasound techniques, suggest indeed that the tendon stiffness is near, or below the critical value.\endnote{\label{DEBATE}C. N. Maganaris and J. P. Paul, ''Tensile properties of the in vivo human gastrocnemius tendon,'' Journal of Biomechanics 32, 1639-1646 (2002); I. D. Loram, C. N. Maganaris, and M. Lakie, ''Human postural sway results form frequent, ballistic bias impulses by soleus and gastrocnemius,'' Journal of Physiology 564.1, 295-311 (2005a), ''Active, non-spring-like muscle movements in human postural sway: how might paradoxical changes in muscle length be produced?,'' Journal of Physiology 564.1, 283-293 (2005b). Ultrasound allows direct measurements of the length changes of the muscle fibers and the tendon.} 

Still, many of the basic questions about balance control remain far from settled. Anyway, we feel that the springy controlled inverted pendulum may be an interesting and simple device for demonstrating an unstable equilibrium system with biological relevance. The effect of the compliant link can be demonstrated by comparing the balancing task with, and without the spring. Instead of manual control one may use a computer controlled actuator pulling the wire, employing various sensors for feedback data (force, inclination angle, spring extension). In fact, muscles are equipped with sensor organs called muscle spindles which basically record muscle length and its rate of change, while the muscle force is gauged by the Golgi tendon organs (GTO) located in the tendons.\endnote{For a standard reference on the human neuro-muscular system see R. M. Enoka, \textit{Neuromechanics of human movement} (Human Kinetics, 2002), 3. ed.} These organs enable the human motor system to act as a feedback control system with delay (due to finite neural conduction speed and processing time). The vestibular system is also important for balance, but during quiet standing the acceleration of the head is normally too small to trigger vestibular reflexes. One set of interesting questions is related to the issue how damage (e.g. due to neural degeneration) might affect balance and the control system, and whether there might be compensatory strategies. Similarly one may investigate using the demonstration device how suppression of feedback data, or a change in the feedback delay, affect its balance control. 

In the following sections we will give a mathematical description of the inverted pendulum system, and a few possible control methods, together with some practical details about the demonstration device. While the use of the pendulum in analyzing walking and running may be well known in physics circles\endnote{\label{AHLBORN}B. K. Ahlborn and R. W. Blake, ''Walking and running at resonance,'' Zoology 105, 165-174 (2002), presents one of the most recent pendulum models for walking and running. (Note that due to some typographic problems there seems to be a lot of missing $\pi$'s in the online paper.) Such pendulum models go at least back to a work by E. Weber and W. Weber, \textit{Mechanik der menschlichen Gehwerkzeuge} (Dietrich, 1836), while the scientific analysis of locomotion began in earnest by G. A. Borelli, the ''father of biomechanics'', in \textit{De motu animalum} (1680) (the same Latin title has been used for a book by Aristotle). The younger brother Wilhelm Weber is by the way known in physics for his work on electromagnetism and for the \textit{Weber}-unit. Boye Ahlborn has also written a delightful textbook, \textit{Zoological physics} (Springer, 2004), with an emphasis on the physical principles underlying animal locomotion.}, the fact that it is also used for analyzing quiet standing may be less well known. As neither the spring-coupled pendulum seems to be very familiar, we find the it justified to present the following detour. The topic also provides a link between basic physics and biology, and it may convince some students that interesting research questions can arise from such deceivingly simple phenomena as quiet standing.

\section{Theory}
\subsection{The human inverted pendulum (HIP)}

We will first consider the HIP model depicted in the Fig. 2. Using the notation of that figure, and applying Newtonian mechanics, we can write the following equations,\endnote{\label{JOHNS}We may note that the same Eq.(\ref{eq:HIP3}) is obtained for the control of an unicycle with the feedback term $N$ given by $mL \ddot y \cos(\theta)$ where $y$ is the position of the wheel in the forward direction. See R. C. Johnson, ''Unicycles and bifurcations,'' Am. J. Phys. 66 (7), 589-592 (1998).}

\begin{eqnarray}
\label{eq:HIP1}
m\ddot y &=& F_y,\\
\label{eq:HIP2}
m\ddot z &=& F_z,\\
\label{eq:HIP3}
I\ddot\theta &=& mgL\sin(\theta) - N,\\
\label{eq:HIP4}
N &=&  u F_z + \zeta F_y.
\end{eqnarray}

Here $I$ denotes the moment of inertia of the body (minus feet)\endnote{The problem of determining the moment of inertia of the (living) human body is an interesting and challenging exercise in itself for students (cadavers are also been used for this purpose but the method is somewhat cumbersome and the samples may not be representative). Two standard references on data and methods for measuring and calculating the momenta of inertia and the centers of mass of body segments are D. A. Winter, \textit{Biomechanics and motor control of human movement} (Wiley, 2005), 3. ed., and V. M. Zatsiorsky, \textit{Kinetics of human motion} (Human Kinetics, 2002). See also I. W. Griffiths, J. Watkins, and D. Sharpe, ''Measuring the moment of inertia of the human body by a rotating platform method,'' Am. J. Phys. 73 (1), 85-92 (2005). Given data on the properties of the body segments one can calculate the moment of inertia of the body with respect to the ankle joints. We may quote a  representative value of $I$ = 66 kg m$^2$, for a male adult with $m$ = 76 kg and $L$ = 0.87 m, from K. Masani et al., ''Importance of body sway velocity information in controlling ankle extensor activities during quiet stance,'' Journal of Neurophysiology 90, 3774 - 3782 (2003).} with respect to the ankle joints, $m$ is the body mass (minus feet - the feet may account for about 3\% of the body mass), $g$ is the gravitational acceleration ($\approx$ 9.81 m/s$^2$), $L$ is the distance from the ankle joints to the center of gravity (COG), $F_y$ and $F_z$ are the components of the ground reaction force (GRF) related to $N$, the torque produced by the plantar flexors counteracting the gravitational torque $mgy$. For small inclinations $\theta$ we can use the approximation $y = L \sin(\theta) \approx L\theta$ in Eq.(\ref{eq:HIP3}). Substituting Eqs.(\ref{eq:HIP1}), (\ref{eq:HIP2}), and (\ref{eq:HIP4}) into Eq.(\ref{eq:HIP3}), and taking into account that\endnote{The vertical ground rection force $F_z$ is not exactly equal to $mg$ during quiet standing. In fact, the heartbeats, and the changing bloodflow (hemodynamics), cause fluctuations in $F_z$ by around 5-8 N, which are however only about one percent of the average value of $F_z$ for an ordinary adult.} $F_z \approx mg$, we get

\begin{equation}
\label{eq:HIP5}
y - u = \left(\frac{\zeta}{g} + \frac{I}{m g L} \right) \ddot y,
\end{equation}

or equivalently

\begin{equation}
\label{eq:HIP6}
\ddot y = \omega_c^2 \cdot (y - u),  
\end{equation}	 

with the characteristic frequency

\begin{equation}
\label{eq:HIP7}
f_c = \frac{\omega_c}{2\pi} = \frac{1}{2 \pi} \sqrt{\frac{g}{\zeta + \frac{I}{m L}}}.  
\end{equation}	 

From Eq.(\ref{eq:HIP4}) we can infer that the muscle torque $N$ is about proportional to the A/P COP coordinate $u$. Indeed, using the above approximations we get $N \approx umg + \zeta m \ddot y$. The term $umg$ is in general much larger than $\zeta m \ddot y$ since $u$ usually varies in the range of 2 - 7 cm while $\ddot y$ may be of the order of ca 1 cm/s$^2$, and $\zeta$ less than 10 cm while $g \approx$ 981 cm/s$^2$. That the A/P COP coordinate $u$ varies with the muscle activity has been verified (by ourselves among others) by measuring Gastrocnemius activity using Electromyography (EMG) during quiet standing, and then comparing the EMG signal with A/P COP.

The HIP model for quiet standing has been tested in several investigations.\endnote{Se for example A. Karlsson and H. Lanshammar, ''Analysis of postural sway strategies using an inverted pendulum model and force plate data,'' Gait \& Posture 5 (3), 198-203 (1997); W. H. Gage et al., ''Kinematic and kinetic validity of the inverted pendulum model in quiet standing,'' Gait \& Posture 19 (2), 124-132 (2004). The validity of the HIP model presupposes that the standing person adopts the so called ankle strategy; that is, controls the balance using the muscles acting over the ankle joints. This comes naturally for most people during quiet standing, but some people, perhaps due to neurogenic or myogenic disorders, may have to keep the balance by moving the hip also. To describe such cases the one-segment HIP model must be replaced with a multi-segment version.} Usually one needs, besides the force plate, elaborate video-systems in order to track the body segment and obtain the resultant COG and its $y$-coordinate. We have, by the way, employed a much simpler system where a thin wire was attached to the person at the waist level (which is close to the COG of a human being), then let to run over a small pulley and finally connected to a lever arm of a rotational optical encoder (resolution of 5000 pulses per revolution). With this arrangement it was easy to measure the backward-forward motion with an accuracy better than 0.1 mm. If one uses this system in combination with the force plate one obtains both the A/P COP coordinate $u$ and the COG coordinate $y$. The HIP model predicts then, by writing Eq.(\ref{eq:HIP6}) in the frequency domain (the ''hat'' denotes the Fourier-transformation of the function), 

\begin{equation}
\label{eq:HIP8}
\hat y(f) = \frac{\hat u(f)}{1 + \left(\frac{f}{f_c} \right)^2},
\end{equation}

that $y$ should be a low-pass filtered version of $u(t)$. This was indeed verified within reasonable limits by computing the low-pass filtered transform of $u$ using Eq.(\ref{eq:HIP8}) with $f_c$ = 1/2 Hz, and then comparing the result with the measured $y$-series.\endnote{A validation study of the spectrum method for calculating A/P COG from A/P COP has been presented by O. Caron, B. Faure, and Y. Breni{\`e}re, ''Estimating the centre of gravity of the body on the basis of the center of pressure in standing posture,'' Journal of Biomechanics 30 (11/12), 1169 - 1171 (1997).} 

\subsection{Feedback control}
\label{FBACK}

Within the above mathematical representation the task of the balance control is to vary the function $u(t)$ (the control function) in Eq.(\ref{eq:HIP6}) such that $y(t)$ remains bounded in a small interval. It is a straightforward exercise to solve for $y(t)$ in term of $u(t)$ as an initial value problem. One can proceed by defining a new variable

\begin{equation}
\label{eq:NEWVAR}
q(t) = y(t) + \frac{1}{\omega_c} \dot y(t),
\end{equation}

which together with 
\begin{equation}
\label{eq:QVERS}
\dot q(t) = \omega_c \cdot (q(t) - u(t))
\end{equation}

is equivalent to Eq.(\ref{eq:HIP6}). Knowing $q(t)$ we can solve for the original variable $y(t)$ from Eq.(\ref{eq:NEWVAR}),

\begin{eqnarray}
\label{eq:YSOL}
y(t) = y(0) \cdot e^{-\omega_c t } + \frac{1}{\omega_c}\int_0^t e^{\omega_c (s - t)}q(s)ds =\nonumber\\
y(0) \cosh(\omega_c t) + \frac{\dot y(0)}{\omega_c} \sinh(\omega_c t) + \omega_c \int_0^t \sinh(\omega_c(s-t)) u(s) ds.
\end{eqnarray}

From this it is apparent that $y(t)$ stays bounded whenever $|q(t)| < C$ for some constant $C$. In fact, the decomposition of Eq.(\ref{eq:HIP6}) into Eqs.(\ref{eq:NEWVAR}) and (\ref{eq:QVERS}) is a basic example of a decomposition into a stable and an unstable manifold of a dynamical system. It makes sense for the control system to address the unstable variable component, since once this is controlled the stable part will take care of itself. We may observe that we also have a solution of the form

\begin{equation}
\label{eq:YSOL2}
y(t) = \frac{\omega_c}{2}\int_{-\infty}^\infty e^{- \omega_c|s - t|} u(s) ds. 
\end{equation}

A direct substitution of Eq.(\ref{eq:YSOL2}) into Eq.(\ref{eq:HIP6}) demonstrates that it is indeed a solution if the integral exists. This may seem like a strange solution because it does not directly depend on the initial values $y(0)$ and $\dot y(0)$, instead it depends on the future values of the control function $u(t)$. When we compute $y(t)$ from the $u(t)$-data using the low-pass filter Eq.(\ref{eq:HIP8}) this will correspond to assuming a solution of the form given by Eq.(\ref{eq:YSOL2}). Indeed, the filter factor $\left( 1 + \left(f / f_c \right)^2 \right)^{-1}$ in Eq.(\ref{eq:HIP8}) is the Fourier transform of the Green's function $G(t) = \frac{\omega_c}{2} \cdot e^{-\omega_c |t|}$ appearing in Eq.(\ref{eq:YSOL2}). Using Eq.(\ref{eq:YSOL2}) we obtain for $q$ the expression

\begin{equation}
\label{eq:QEQU2}
q(t) = \omega_c \int_t^{\infty} e^{- \omega_c (s - t)} u(s) ds,
\end{equation}

which agrees with the causal solution if\endnote{Professor Olof Staffans (Math. dept., The Abo Akademi University) pointed out to me that this property is linked to the concept of ''exponential dichotomy'' in the field of Dynamical Systems. Those who have encountered advanced/retarded solutions in electrodynamics and the action-at-a-distance formulations (e.g. the Wheeler-Feynman theory) may see a connection here too; for a review see F. Hoyle and J. V. Narlikar, ''Cosmology and action-at-a-distance electrodynamics,'' Rev. Mod. Phys. 67, 113-155 (1995), which has also appeared in a book form as \textit{Lectures on cosmology and action-at-a-distance electrodynamics} (World Scientific, 1996).

We may note that there is a simple discrete analogy to the case (\ref{eq:QEQU2}) in the form an equation $x_{k+1} = a \cdot x_k - u_k$ with $a > 1$. Suppose the control function $u_k$ is chosen such that $x_k$ remains bounded, then $x_k$ may be expressed in terms of the future $u$-values by  

\[
x_k = \frac{1}{a} \sum_{j \geq k} u_j a^{k-j} = \frac{u_k}{a} + \frac{u_{k+1}}{a^2} + \dots \,,
\]

which follows by developing $x_k = u_k/a + x_{k+1}/a$ and assuming that $x_N/a^N \rightarrow 0$ as $N \rightarrow \infty$. Thus, in this case $x_k$ can be determined without knowing the initial values.

}  

\begin{equation}
\label{eq:CONDYSOL2}
q(t)  e^{- \omega_c t} \rightarrow 0 \quad \mbox{as} \quad t \rightarrow \infty.
\end{equation}

We may thus consider it justified to use Eq.(\ref{eq:YSOL2}), or rather its Fourier version, when we are concerned with bounded motion (no falling).


In view of Eq.(\ref{eq:QVERS}) one may design a threshold controller\endnote{A similar feedback function for balance has been considered by C. W. Eurich and J. G. Milton, ''Noise-induced transitions in human postural sway,'' Phys. Rev. E 54 (2), 6681-6684 (1996). However, they start from the equation for a damped inverted pendulum (we replacw here $\sin(\theta)$ by $\theta$), $m R^2 \ddot \theta(t) + \gamma \dot \theta(t) - mgR \theta(t) = f(\theta(t -\tau))$, and argue that the system is over damped and that the $\ddot \theta$-term may therefore be dropped in comparison, thus arriving at a first order differential equation. To assume such a large friction coefficient $\gamma$ for the ankle joint appears nonphysiological. (The static and dynamic friction coefficients for synovial joints are about $\mu_s$ = 0.01 and $\mu_k$ = 0.003, to be compared with $\mu_k \approx$ 0.05 for lubricated ball bearings.) If the motion seems like being heavily damped this might be the result of an active control, and it is how this can be achieved which one has to try to explain in the first place. Since we use the variable $q$ of Eq.(\ref{eq:NEWVAR}) we can employ a similar feedback control, but now in term of $q$, without the need of recourse to the hypothesis of over damping. In a recent paper, A. L. Hof, M. G. Gazendam, and W. E. Sinke, ''The condition of dynamic stability,'' Journal of Biomechanics 38 (1), 1-8 (2005), the authors introduce the combination $q$, which they call ''the extrapolated center of mass position (XcoM)'', in a biomechnical analysis the ''base of support'' (BoS).

Milton has been involved in another interesting study of the inverted pendulum, namely in investigating the balancing of a stick on the tip of a finger; see J. L. Cabrera and J. G. Milton, ''On-off intermittency in human balancing task,'' Phys. Rev. Lett. 89 (15), 158702 (2002). In this case the pendulum is controlled by moving the pivot point (finger). Based on their data Cabrera and Milton concluded that the time series of the tilt angle exhibited characteristics of the so called L{\'e}vy-flight. For a critical review see F. Borg, physics/0411138. An interesting issue is whether one could find some similarities between the dynamics of quiet standing and stick balancing.} which starts to pull on the pendulum whenever $q$ crosses a threshold value $q_{th}$ (''bang''-control); that is, the feedback is of the form $u(t)=f(q(t-\tau))$, with a delay $\tau$ included,

\begin{equation}
\label{eq:FBF}
f(q) = \left\{ \begin{array}{ll}
0 & \mbox{if} \quad q \leq q_{th} + \epsilon_2\\
C + \epsilon_1 & \mbox{otherwise} 
\end{array} \right..
\end{equation}
 
Here the parameter $C$ determines the strength of the feedback force, while $\epsilon_i$ represent additional stochastic elements (''noise''). Thus, when $q$ exceeds a threshold $q_{th}$ plus a random fluctuation, a controlling force $C + \epsilon_1$ will act with a delay $\tau$. The equation of motion becomes,

\begin{equation}
\label{eq:FBACK}
\dot q(t) = \omega_c q(t) - \omega_c f(q(t - \tau)),
\end{equation}

and it is apparent that this system may sustain an oscillatory motion (for a simulation see Fig. 3). Indeed, let's first neglect the ''noise'' terms, then if we start from $q(0) < q_{th}$ we will have an exponential increase $q(t) = q(0) \cdot \exp(\omega_c t)$ until $q(t - \tau)$ reaches the threshold value $q_{th}$. Then, if $C > q_{th} \cdot \exp(\omega_c \tau)$, the feedback force will reverse the motion and the force  persists until $q(t - \tau)$ crosses the threshold again from the other side, and so on. The system thus settles into a periodic motion whose period can be calculated to be

\begin{equation}
\label{eq:PERIOD}
T = 2\tau´+ \frac{1}{\omega_c} \ln\left( \frac{C - q_{th}}{C - e^{\omega_c \tau} \cdot q_{th}} \right) + \frac{1}{\omega_c} \ln\left(\frac{q_{th}}{q_{th} - \left( C - q_{th}\right) \left(e^{\omega_c \tau} - 1 \right)} \right),
\end{equation}

if the following requirement for bounded motion is satisfied,

\begin{equation}
\label{eq:COND1}
C_{max} \equiv q_{th} \cdot \left(1 - e^{-\omega_c \tau} \right)^{-1} > C > C_{min} \equiv q_{th} \cdot e^{\omega_c \tau}.
\end{equation}

Computer simulations of Eq.(\ref{eq:FBACK}) with added (not too large) noise still produce bounded oscillations. The oscillating case may be regarded as an ''attractor'' of the postural control system. From Eq.(\ref{eq:COND1}) we see that $C_{max} / C_{min} = \left(e^{\omega_c \tau} - 1 \right)^{-1}$, and because this ratio must be larger than 1 for bounded motion, there is an upper stability limit for the delay $\tau$ set by

\begin{equation}
\label{eq:LIMIT1}
\tau_{max} = \frac{\ln(2)}{\omega_c},
\end{equation}

yielding $\tau_{max} \approx 230 \;\mbox{ms}$ for a typical adult value $\omega_c \approx 3 \;\mbox{s}^{-1}$. This conclusion is of course only valid with respect to this particular feedback model. Still, the human neuro-motor postural control system operates with feedback delays in the range from 40 ms (myotatic stretch reflexes), and 100 ms (''programmed'', automatic postural responses), to ca 150 ms (voluntary postural movements), depending of the pathway (spinal pathway; brain stem and subcortical pathway; cortical pathway). So, in this sense the model is within physiological limits. The model also mimics the physiological situation in that it only uses a pulling feedback force which corresponds to the fact that during quiet standing only the plantar flexors are active. Furthermore, the ansatz for the feedback force Eq.(\ref{eq:FBF}) implies that the neuro-muscular system employs the information about muscle length plus its rate of change for the control of balance, in the form of the combination $q(t) = y(t) + {\dot y(t)} / {\omega_c}$. Physiologically this is possible since the muscle spindle confers information about the muscle length ($x_1$) and its rate of change ($\dot x_1$). True, $y$ is not directly proportional to the muscle length $x_1$, but to the total muscle-tendon length $l$ ($l = x_1 + x_2$, where $x_2$ is the tendon length). However, if $x_1$ is known from spindle data, and the neuro-muscular system can infer the tendon length ($x_2$) from the force ($F$) data provided by the Golgi tendon organ (GTO) using some learned empirical tendon force-length relation, $F = F(x_2)$, then an estimate of the total length $l$ (and consequently of $y$) will be available for the feedback control. 

\subsection{Instability through compliance}

By simple geometry the Gastrocnemius muscle-tendon lengthens, for an adult, by about 1 mm per degree of forward inclination. A study\endnote{D. W. Grieve, S. Pheasant, and P. R. Cavanagh, ''Prediction of Gastrocnemius length from knee and ankle posture,'' in \textit{Biomechanics VI-A, Proceedings of the sixth international congress of biomechanics, Copenhagen´, Denmark}, edited by E. Asmussen and K. J{\o}rgensen (University Park Press, 1978), pp. 405-412. The limbs in the study were from people aged 60 years plus.} of 8 cadaveric limbs yielded the relationship,

\begin{equation}
\label{eq:LENGTH}
100 \cdot \frac {\Delta l}{l} = -22.18 + 0.30 \cdot \vartheta - 0.00061 \cdot \vartheta^2
\end{equation}  

for the length change in percent segment length as a function of the ankle angle $\vartheta$ expressed in degrees. The angle $\vartheta$ = 86.8$^\circ$ corresponds to a 3.2$^\circ$ forward leaning position (and to $y \approx$ 5 cm) and at this point we get from Eq.(\ref{eq:LENGTH}) that ${\partial \Delta l}/{\partial \vartheta} \approx l \cdot (0.196\%)$; thus, for a shank length of 400 mm the length change becomes 400 mm $\times$ 0.196/100 $\approx$ 0.8 mm per degree. This has interesting consequences when we consider the tendon properties. As the muscle and the tendon are in series the total length is $l = x_1 + x_2$, where $x_1$ is the muscle length and $x_2$ is the tendon length. The elastic properties of the tendon is determined by the relation between its elongation ($\Delta x_2$) and the load ($\Delta F$). Maganaris and Paul (see note \ref{DEBATE}) have, among others, tried to measure the tendon elongation as function of the load using ultrasound viewing in vivo. The result is that the tendon behaves as a nonlinear spring. Mapping data from their published graph (based on data from 8 young male adults) and fitting a 2nd order polynomial gives the relationship ($\Delta F$ in units of N, and $\Delta x_2$ in units of mm),

\begin{equation}
\label{eq:TENDON}
\Delta F = 39.1 \cdot \Delta x_2 + 3.4 \cdot {\Delta x_2}^2.
\end{equation}  

This covers a force range of 0 - 870 N and an elongation range of 0 - 11 mm. During unloading the force $\Delta F$ was about 18\% larger than during loading for the same elongation (hysteresis). Eq.(\ref{eq:TENDON}) is to be regarded mainly as an illustrative example, but the nonlinear behaviour of the tendon is a general feature. From Eq.(\ref{eq:TENDON}) we can calculate the tendon stiffness $K$ by,

\begin{equation}
\label{eq:STIFF}
K = \frac{\partial F}{\partial \Delta x_2},
\end{equation}  

which yields e.g. $K$ = 81 N/mm when $\Delta x_2$ = 6.2 mm. Using Eqs. (\ref{eq:STIFF}), (\ref{eq:TENDON}), and (\ref{eq:LENGTH}), one may estimate the torque $r \times \Delta F$ generated by tendon for a given elongation,  assuming a moment length $r$ = 0.05 m vis-\`a-vis the heel. Thus, suppose we have a person with $m$ = 76 kg, $L$ = 0.9 m, who leans forward by 0.05 m in term of COG $y$ (an inclination around $\theta$ = 3.2$^\circ$ ). The weight will then be 373 N per leg on the average, corresponding to an elongation $\Delta x_2$ = 6.2 mm and a stiffness $K$ = 81 N/mm. Assuming that the tendon lengthens by 1 mm per 1 degree of inclination, it follows that both (left and right leg) Achilles tendons together would generate a torque of 464 Nm/rad ($ = 2 \times r \times {\partial \Delta F}/{\partial \Delta \theta} = r \times {\partial \Delta F}/{\partial \Delta x_2} \times {\partial \Delta x_2}/{\partial \Delta \theta}$ = 2 $\times$ 81 $\times$ 180/$\pi$ Nm/rad, the last factor coming from 1$^\circ$ = $\pi$/180 radians), to be compared with the gravitational ''stiffness'' $mgL$ = 671 Nm/rad. That is (see Fig. 4), if the muscle locks its length $x_1$ and leaves it to the tendon to rebound from any forward disturbance ($\Delta \theta > 0$), then the torque generated by the tendon will be overcome by gravity and the person topples over (the resulting torque being $\Delta T_{tot}$ = (671 - 464) Nm/rad $\times$ $\Delta \theta$ for small disturbances $\Delta \theta$). Conversely, for a backward disturbance ($\Delta \theta < 0$) the tendon will win over gravity ($\Delta T_{tot} < 0$) and the person falls on his/her back. The implication is that the muscle must actively change its length in response to disturbances so that the ''effective stiffness'' of the muscle-tendon system is larger than the gravitational stiffness. 

For a simple model of how the ''effective stiffness'' can be affected, assume that the muscle manages to keep the proportion of length change of the muscle, $\Delta x_1 = x_1 - x_1^0$, and the tendon, $\Delta x_2 = x_2 - x_2^0$, constant; that is,

\begin{equation}
\Delta x_2 = -\gamma \cdot \Delta x_1. 
\end{equation}

The change in the total length becomes $\Delta l = \left(1 - 1/\gamma \right) \Delta x_2$. Thus, the force exerted by of the tendon can be written $K {\Delta x_2}$ = $K\left(\gamma/(\gamma - 1)\right){\Delta l}$, which implies that the ''effective'' muscle-tendon stiffness is 

\begin{equation}
\label{eq:EFFSTIFF}
K_{eff} = \left( \frac{\gamma}{\gamma - 1} \right) \cdot K.
\end{equation}  

Therefore, if the muscle contracts half as fast as the tendon lengthens ($\gamma$ = 2) then the effective muscle-tendon stiffness will be twice as large as the tendon stiffness.

The threshold feedback model discussed in section (\ref{FBACK}) did not directly relate to the intrinsic instability caused by compliance, since the feedback control was formulated in term of a feedback force not caring about how this force is transmitted (such as by a springy link). However, the feedback force must be related to the spring elongation by (linear example and not counting hysteresis)

\begin{equation}
\label{FBSPRING}  
f(q(t - \tau)) = K \left( x_2(t) - x_2^0 \right),
\end{equation}

where $x_2^0$ is the tendon length at the ''operating'' point. This adds a compatibility condition for the feedback control since $x_2$ has a restricted range. Physiologically the ''bang''-character of the feedback control Eq.(\ref{eq:FBF}) would be rather odd since, as the force switches between 0 and $C$, the tendon length would switch between $x_2^0$ and $x_2^0 + C/K$. We can hardly expect such a discrete behaviour in reality. For instance, the muscle contracts with a finite velocity. Yet, if we look at the level of muscle cells (fibers), then we have more or less an on-off behaviour. The fibers of the muscle are organized in motor units (MU), each controlled by a single motor nerve, such that a MU is either on or off. The total force of the muscle depends on the number of motor units activated. Thus, in a more refined feedback model the number of MUs activated could be a (probabilistic) function of $q$. The stochastic terms in Eq.(\ref{eq:FBF}) partly reflects such an approach through the fluctuations in the threshold level and the force. 

\subsection{PID-control}

The most common approach, in an engineering context at least, is to assume a PID-type feedback control in which the feedback torque $N$ (Eq.(\ref{eq:HIP4})) is proportional to deviation (plus its derivative and its integral) from the desired position, as for instance described by Masani et alii,\endnote{\label{MASANI05} K. Masani, A. H. Vette, and M. R. Popovic, ''Controlling balance during quiet standing: Proportional and derivative controller generates preceding motor command to body sway position observed in experiments,'' Gait \& Posture (article in press, online). The controller discussed in this paper is really a PD-controller only since the integration (I-) term is not used. For a reference on control theory in the physiological context and with Matlab codes, see M. C. K. Khoo, \textit{Physiological control systems. Analysis, simulation, and estimation} (IEEE Press, 2000).}

\begin{equation}
\label{eq:PID}
N(t) = -K_D \, \dot \theta(t - \tau) - K_P \, \theta(t - \tau). 
\end{equation}

The authors decompose the delay as $\tau = \tau_F + \tau_M + \tau_E$ where $\tau_F$ is termed ''feedback delay'' assumed to be ca 40 ms, $\tau_M$ is the ''motor command time delay'' for which they used 135 ms, and $\tau_E$ is the ''electromechanical delay'' estimated to be around 10 ms. The parameter $K_P$ in Eq.(\ref{eq:PID}) is not the (passive) muscle-tendon stiffness constant but describes a gain of the active muscular feedback system. Whether the resulting equation of motion has stable solutions can be investigated by inserting $\theta(t) \propto e^{\lambda t}$ which yields (assuming $\sin (\theta) \approx \theta$),

\begin{equation}
\label{eq:CHAR}
I \lambda^2 + K_D \lambda \, e^{- \lambda \tau} + K_P \, e^{- \lambda \tau} - mgL = 0.
\end{equation}

If the real part $\Re(\lambda)$ of its solutions $\lambda$ satisfies $\Re(\lambda) < 0$ then stability is ensured. For instance, using $m$ = 76 kg, $I$ = 66 Nm s$^2$, $L$ = 0.87 m, $K_P$ = 750 Nm/rad, $K_D$ = 350 Nm s/rad, and $\tau$ = 185 ms we get the numerical solution $\lambda \approx$ -0.49 s$^{-1}$ to Eq.(\ref{eq:CHAR}) thus implying a stable case. If we replace the derivative term in Eq.(\ref{eq:PID}) with an integrated average, such as $\int_{t - \tau}^{t} \dot \theta (s - \tau) ds$, we can obtain a proportional minus delay (PMD) controller\endnote{I. H. Suh and Z. Bien, ''Proportional minus delay controller,'' IEEE Transaction on Automatic Control, AC-24 (2) 370-2 (1979).} of the special form,

\begin{equation}
\label{eq:PMD1}
N(t) =  A \, \theta(t - \tau) + B \, \theta(t - 2\tau).
\end{equation}

In fact, Atay (1999)\endnote{F. M. Atay, ''Balancing the inverted pendulum using position feedback,'' Appl. Math. Lett. 12 (5) 51-56 (1999).} has shown that the system

\begin{equation}
\label{eq:PMD2}
\ddot x(t) + k x(t) = a x(t - 1) + b x(t - 2)
\end{equation}

can be stabilized in case of $k < 0$ (inverted pendulum) for special choices of $a$ and $b$. Eq.(\ref{eq:PMD2}) can be related to our case if we rescale time as $t \rightarrow t/\tau$, set  $k = - (\omega \tau)^2$, $a = A \tau^2$, and $b = B \tau^2$. The characteristic equation for Eq.(\ref{eq:PMD2})
is

\begin{equation}
\label{eq:PMD3}
\lambda^2 + k - a e^{-\lambda} - b e^{-2\lambda} = 0,
\end{equation}

and Atay proves that, for $k < 0$, we have solutions $\Re(\lambda) < 0$ if and only if, 
  
\begin{eqnarray}
&(a)& \quad k > -1,\\
&(b)& \quad k < b < \left(\frac{\pi}{2}\right)^2 - k,\\
&(c)& \quad -2 b \cos \sqrt{k + b} < a < k - b. 
\end{eqnarray}  

Condition $(a)$ for instance means that we must have $\tau < 1/\omega_c \approx$ 300 ms using the typical value $\omega_c \approx$ 3 s$^{-1}$. Thus, this special PMD-controller seems to be able to ensure stability with a bit longer delay than the ''bang''-control with the limit given by Eq.(\ref{eq:LIMIT1}). A direct numerical evaluation shows that, using for example $\omega_c$ = 3 s$^{-1}$, $a$ = -0.9 and $b$ = 0.6, we obtain the root $\lambda \approx$ -0.289.  

The point of mentioning the PMD-control model is that it shows that knowledge of the point-derivative is not necessary for stabilization, but that time-shifted copies of the position might do instead. Theoretically there are apparently a large number of control methods available for quiet standing. Which ones might be realized physiologically? Again, a common engineering approach is to study the class of linear models (ARMAX, Autoregressive-Moving Average process) and try to find the parameter values by fitting the  model to experimental data (System Identification), this is especially the method followed by Peterka.\endnote{R. J. Peterka, ''Simplifying the complexities of maintaining balance,'' IEEE Engineering in Medicine and Biology Magazine, 63-68 (March/April 2003).} Yet we would like to know whether balancing implements some sort of an optimal strategy, considering the requirement of robustness in a noisy environment and the need to economize with the muscle energy. For instance, ''bang-bang'' type controlls may result when one tries to minimize the time for going from one position to another. Optimization principles have been considered for voluntary movements, such as walking and reaching, but quiet standing seems to be harder to adapt to such a procedure. One objective may be to try to keep the swayings below the vestibular trigger level and such that no extra limb or hip movements are required. Given any control model one can ask which parameter regime will keep the system in a reasonable physiological range. Thus, in the ''bang''-model, the proper average feedback force $C$ must be somewhere in the interval given by Eq.(\ref{eq:COND1}) such that the projection of the COG will stay within ca 10-20 mm of the middle of feet (defining the threshold point) in order not to provoke extra stabilizing measures too often. Fig. 5 shows the stable range for $C$ as a function of the delay $\tau$. The amount of ''noise'' and the size of the delay $\tau$ will constrain the choices of $C$. 

Finally one may wonder why nature uses compliant tendons, which are the essential advantages? They may enable a smoother control, protect muscles at sudden pulls, and store potential energy in jumps and cyclic movements.\endnote{R. McN. Alexander, ''Storage and release of elastic energy in the locomotor system and the stretch-shortening cycles,'' in B. M. Nigg, B. R. Macintosh, and J. Meister (eds.), \textit{Biomechanics and biology of movement} (Human Kinetics, 2000) 19-29.}

\subsection{Bifurcations}

One question of principal interest with regards to the inverted pendulum control is whether there is sort of bifurcation phenomena with regards to, say, the delay parameter\endnote{\label{YAO} The paper W. Yao, P. Yu, and C. Essex, ''Delayed stochastic differential model for quiet standing,'' Phys. Rev. E 63, 021902 (2001), presents a delay-bifurcation analysis based on the model of Eurich and Milton referred to above. Note that the authors give an erroneous mathematical description of the inverted pendulum - among other things they identify COP with COG - but if we use instead of their $x$ our $q$, as in Eq.(\ref{eq:FBACK}), then one can still apply their results. A bifurcation analysis in term of reflex gain and delay has been presented by B. W. Verdaasdonk et al., ''Bifurcation and stability analysis in musculo-skeletal systems: a study of human stance,'' Biological Cybernetics 91, 48-62 (2004). This study, however, assumes that we have an ''ankle stiffness'' generated by a coactivation of Gastrocnemius and Tibialis anterior (TA, which pulls forward), while in fact TA is mostly silent during quiet standing. For an introduction to bifurcations in biological systems see A. Beuter et al. (eds.), \textit{Nonlinear dynamics in physiology and medicine} (Springer, 2003), and especially the chapter by M. R. Guevara. } $\tau$ or the spring constant $K$. We have argued that for a subcritical spring constant a robust length-locking control is no longer possible and that a phasic control becomes necessary. Thus, if we consider the simple control model which keeps the muscle length constant, we will have a ''bifurcation'' in the stiffness parameter $K_\theta$ (in units of Nm/rad, stiffness as Nm/rad related to stiffness as Nm/mm by $K_\theta \Delta \theta = K \Delta x_2$) when it crosses the critical value of $mgL$. The same is true also for the PID-control Eq.(\ref{eq:PID}) for small delays $\tau$. The general meaning of a bifurcation point $\mu_0$ for a dynamical system depending on a parameter $\mu$,

\begin{eqnarray}
\label{eq:DYNS}
\dot x = f(\mu, x),\nonumber \\
\mu \in (a , b),
\end{eqnarray}

is that the topology of the phase portrait of Eq.(\ref{eq:DYNS}) changes when $\mu$ crosses the value $\mu_0$. A simple example is obtained by setting $f(\mu, x) = \mu - x^2$ which for $\mu < 0$ has no equilibrium point, while for $\mu$ = 0 there is exactly one equilibrium point $x$ = 0 which for $\mu > 0$ ''bifurcates'' into a stable $(x = \sqrt{\mu})$ and an unstable $(x = -\sqrt{\mu})$ equilibrium point (''sink'' resp. ''source''). For a delay equation, like the simple linear feedback model (a Stochastic Delay Differential Equation studied also by Yao et al.$^{\ref{YAO}}$)

\begin{equation}
\label{eq:FBL}
\dot q(t) = \omega_c q(t) - \beta q(t - \tau) + \epsilon\,,
\end{equation}

it may not be that obvious what ''bifurcation'' might mean because we do not have the phase portrait as in the case Eq.(\ref{eq:DYNS}). (In Eq.(\ref{eq:FBL}) $\beta$ characterizes the magnitude of the feedback force and $\epsilon$ represents a ''noise'' contribution.) In this case $\tau$ is a bifurcation parameter in the sense that the stability properties of the system may change as $\tau$ changes. Thus, for $\tau$ = 0 it is apparent that the system Eq.(\ref{eq:FBL}) is stable only if $\beta > \omega_c$. For nonzero $\tau$ we can again study behaviour of the system using the characteristic equation, neglecting the noise,

\begin{equation}
\label{eq:CHAR2}
\lambda - \omega_c + \beta e^{-\lambda \tau} = 0.
\end{equation}

If we assume $\lambda \tau$ to be small and set $e^{-\lambda \tau} \approx 1 - \lambda \tau$ then we obtain from Eq.(\ref{eq:CHAR2}) 

\[
\lambda \approx \frac{\omega_c - \beta}{1 - \beta \tau}
\]
 
which indeed yields a negative value for $\lambda$ in case $\beta > \omega_c$. It also suggests that instability may enter the picture when $\tau$ approaches $1/\beta$. Using the variable $\mu$ = $\lambda \tau$ and decomposing it into the real and imaginary parts, $\mu = x + iy$, we can write Eq.(\ref{eq:CHAR2}) as,

\begin{eqnarray}
\label{eq:CHAR2a}
x + \beta \tau e^{-x} \cos y - \omega_c \tau &=& 0,\\
\label{eq:CHAR2b}
y - \beta \tau e^{-x} \sin y &=& 0.
\end{eqnarray}  
   
From Eq.(\ref{eq:CHAR2b}) it follows $(y \neq 0)$ that $e^{x} = \beta \tau |\sin y/y| \leq \beta \tau$, whence $\beta \tau < 1$ implies that $x < 0$; i.e., stability. Typically we have for $\beta \tau < 1$ two roots on the $x$-axis with $x < 0$. When $\beta \tau$ approaches 1 the roots merge and then split off from the $x$-axis. A further analysis shows that this happens for $\tau = \tau^{\star}$ determined by the equation $\omega_c \tau^{\star} = 1 - \ln \left(1/(\beta \tau^{\star})\right)$ in the interval $(1/(\beta e), 1/\beta)$. One may investigate how $x$ changes at the point $x$ = 0 with respect to $\tau$ by calculating $\dot x = dx/d\tau$ from Eqs.(\ref{eq:CHAR2a}) and (\ref{eq:CHAR2b}), which yields,

\begin{equation}
\label{eq:XNULL}
\dot x \left(1 - \omega_c \tau \right)^2 = \left(\beta^2 - \omega_c^2 \right) \tau \quad \mbox{at} \quad x = 0.
\end{equation}
 
From this it follows, when $\beta > \omega_c$, that $\dot x > 0$ and $x$ thus becomes positive when $\tau$ increases; i.e., an instability emerges. These features can be nicely studied by plotting the contour/surface map of

\[
F(x, y) = |\mu + \beta \tau e^{-\mu} - \omega_c \tau | \quad ( \mu = x + iy),
\]
 
and varying the parameters. (For plotting purposes it may better to use the logarithm $\ln F(x, y)$ instead of $F(x, y)$.) 

Besides bifurcations associated with the delay $\tau$ and the stiffness $K$  we may also mention a sort of bifurcation associated with disturbance; e.g., during quiet standing the force plate is suddenly translated with a given velocity $v$. For small disturbances the person is able to maintain the balance using the ''ankle strategy'' alone, but when the disturbance reaches a critical range new kinds of strategies become necessary (moving the limbs, moving the hip, etc).

\section{{\label{DEVICE}} A demonstration device}

In order to actually feel how a springy controlled inverted pendulum behaves one of course has to build one, and then try to balance it by operating it manually. I use to challenge visitors by letting them try their hands on a device we have next to the office. None has mastered it right off. Indeed, it takes a bit of practice to learn how to intermittently pull and release the wire so that the pendulum will sway around an average forward leaning position and does not topple over. By using springs with varying degree of stiffness one can compare how stiffness affects the task. The construction of a demonstration device is straightforward (see Fig. 6). In our case we used a steel shaft supported by bearings, and attached to a foundation made of a heavy metallic plate. For the pivot one could also use some (discarded) electric motor (kW-size), which gives stable support when bolted to a foundation, and has good bearings. One can also build miniature versions. There are several ways to add computer control to the system, the most challenging part being to design the mechanical actuator, and to equip the system with reliable sensors for generating feedback data. Force transducer, rotational encoders, and potentiometers are the most obvious sensor choices. It may be too tough for inclinometers that are based on accelerometers because they pick up vibrations which mess up the signal. Yet, the intention is not to boost the input data quality, the task is rather to try to get the system working with a minimum level of input data. Thus, one may start with a high grade input data and then add noise, delays, etc, and check how the control algorithm copes with the situation. When it comes to actuators the choice in an industrial context would be a linear motor. A cheaper alternative that we have tested is to use a pneumatic cylinder controlled by the computer via solid state relays connected to magnetic air valves. Our data acquisition system was based on the multipurpose 16 bit A/D card NI PCI-6036E (National Instruments), but any equivalent will do, the sample rate is not critical. The programs were written using NI Labwindows CVI, a C-based tool which may appeal to those who are used to programming in C/C++. (One can also program the NI-cards directly using any standard C/C++-compiler by employing the libraries that come with the cards.) In the simplest version one can use the Timer-function for interrupting, reading data and generating outputs. Set at maximum speed the update intervals on our system were on the average around 54 ms. If one feeds the potentiometer (or other transducer) with current from the computer one will get extra disturbances (e.g. spikes when the hard disk turns on/off), so it may be advisable to use an external regulated source or a battery.

\section{Conclusion}

In this paper we have highlighted the versatility of the pendulum as a model in science by describing yet another application, in this case to the modelling of human standing, which is one of the classical problems of biomechanics. We have also discussed some of the intriguing physical aspects of the model such as instability and delayed control which have become a major research topic in physics in recent years.

\section{Acknowledgments}

Thanks to CEO Mats Manderbacka, HUR Co. (\url{http://www.hur.fi}), for support and help with the parts for the demonstration device. Part of this work is based on a project (\textsc{Bema}+\textsc{Bisoni}) on Biosignals sponsored by the Finnish national technology agency (\textsc{Tekes}). I am indebted to Prof. Ismo Hakala for the opportunity to work at the Chydenius Institute (Jyv{\"a}skyl{\"a} University). 

\theendnotes




\section{Figures and captions}

\begin{description}

\item[Fig1] The human inverted pendulum model (HIP). The pendulum is leaning forward by an angle $\theta$ while being supported by the plantar flexor muscles Gastrocnemnius (GA) and Soleus, of which GA is indicated in the figure. These muscles are attached to the heel bone via the Achilles tendon. The moment length $r$ with respect to the ankle joint is about 5 cm for adults. The ankle joint is taken as the origin of the $(y,z)$ coordinate system. The $y$-axis represents the forward (anterior) direction. If the person stands on a rectangular force plate, then one can measure the center of pressure (COP) as explained in the text. In the figure we have shown the force component $F_1 + F_2$ due the rear transducers numbered 1 and 2, and the force component $F_3 + F_4$ due to the front transducers 3 and 4. The COP-component in the $y$-direction (A/P COP) is denoted by $u$. 

\item[Fig2] The graph shows the time series (called stabilogram) of the A/P COP coordinate minus its average, $u - \langle u \rangle$, during quiet standing. A lot of research has gone into attempts to extract meaningful information from this apparently random curve and the corresponding one for lateral sways.

\item[Fig3] The graphs show a result of simulating the feedback case Eq.(\ref{eq:FBF}). The ''saw'' line represents the unperturbed data $q(t)$ with noise $\epsilon_i$ set to zero. The thick line shows the perturbed result in terms of the COG-coordinate $y$. For the parameters we used $C$ = 80; $q_{th}$ = 40; $\omega_c$ = 3.142 s$^{-1}$ ($f_c$ = 0.5 Hz); delay $\tau$ = $0.5 \times 1/\omega_c$ = 159 ms; integration time step $\Delta t$ = $0.01/\omega_c$ = 3.18 ms. For ''noise'' we used uniformly distributed random numbers in the interval [-10, 10] ($\epsilon_1$) and [-25, 25] ($\epsilon_2$). 

\item[Fig4] A schematic illustration of the case when the muscle length $x_1$ is fixed and only the tendon changes length. The torques excerted by the spring and gravity are equal at the equilibrium point EP corresponding to the inclination $\theta_{EP}$. For a subcritical spring constant $K$ we have the situation depicted by the figure. For inclination $\theta > \theta_{EP}$ gravity wins causing the pendulum to fall forwards, and for $\theta < \theta_{EP}$ the spring wins causing the pendulum to fall backwards, and the equlibrium point is thus an unstable one.

\item[Fig5] The diagram shows the stability area bounded by $C_{max}$ and $C_{min}$ for the feedback parameter $C$ as a function of the delay $\tau$ for the ''bang''-control Eq.(\ref{eq:FBF}), neglecting noise. The illustration is for the case $\omega_c \approx$ 3.14 s$^{-1}$. For small delays the range of allowed $C$ grows rapidly. When $\tau$ increases the stabile $C$-range decreases and evaporates at $\tau_{max}$ = $\ln (2) / \omega_c \approx$ 221 ms.    

\item[Fig6] A device for demonstrating the springy control. It consists of a shaft pivoted by bearings (at Pi) and controlled by the force $F$ which is transmitted via a wire, running over a pulley (P), in series with a spring. When the spring constant is below a given critical value it is no longer possible to lock the pendulum into a stable tilted position. If the spring constant is above the critical value then it is possible to keep the pendulum steady by simply keeping the wire fixed. For subcritical values one has to intermittently pull and release the wire in order to stop the pendulum from toppling over. Our device has a shaft length of 1.02 m and its natural frequency of the pendulum can thus be estimated to be $f_c \approx$ 0.86 Hz. The pneumatic cylinder has a radius of 10 mm, while the piston has a radius of 4 mm, so that when operating at a pressure of 5 bar the corresponding pulling force will be about 130 N. By adjusting the pressure one can thus control the magnitude of the feedback force too. The system managed to keep the balance using a spring which elongated from 180 mm to 270 mm when loaded with 5.85 kg. For feedback control we used the threshold control and feedback variable $\theta + B \dot \theta$ where $B$ is a parameter (''velocity gain'') that could be tuned in real time through the control program. The other adjustable parameter was the threshold level. 


\end{description}

\newpage

\begin{figure}
\includegraphics[scale=0.6]{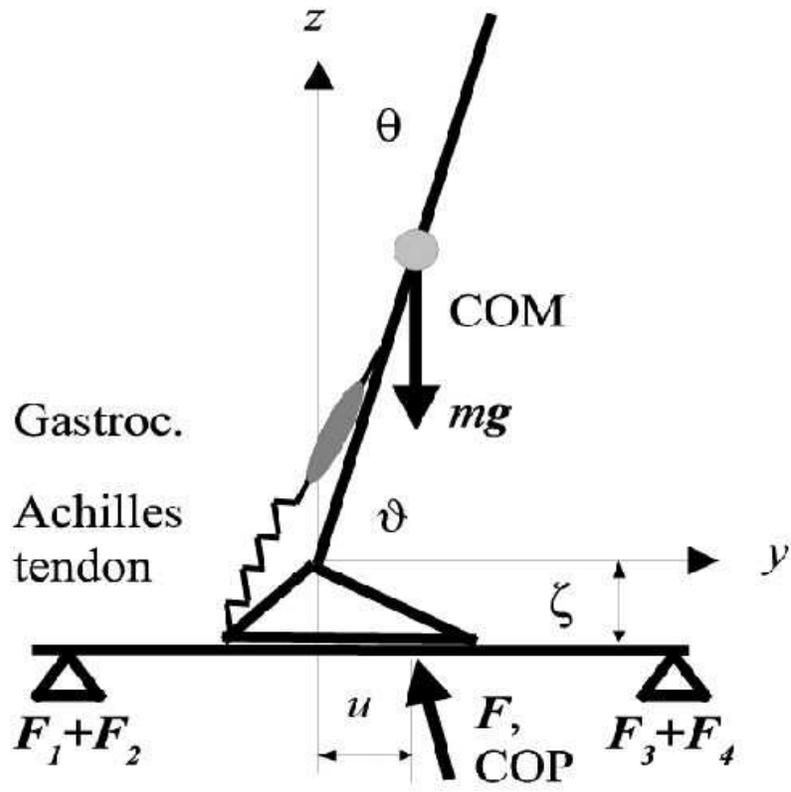}
\caption{The basic inverted pendulum model of standing.}
\label{FIG1}
\end{figure}

\begin{figure}
\includegraphics[scale=0.6,angle=180]{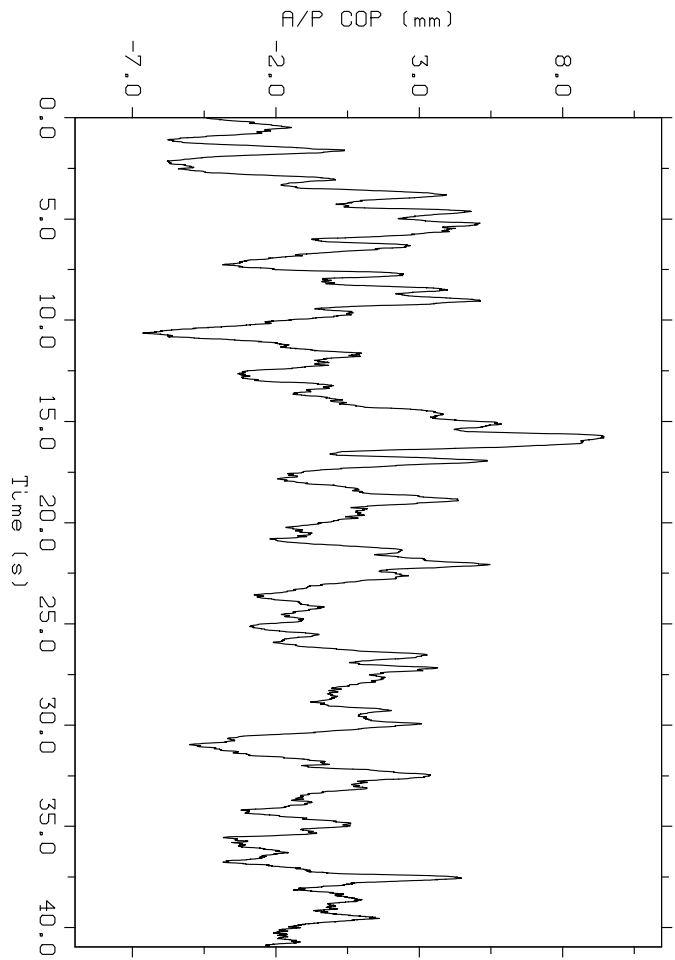}
\caption{An example of the measured forward-backward sway in term of the center of pressure (COP).}
\label{FIG2}
\end{figure}

\begin{figure}
\includegraphics[scale=0.6,angle=180]{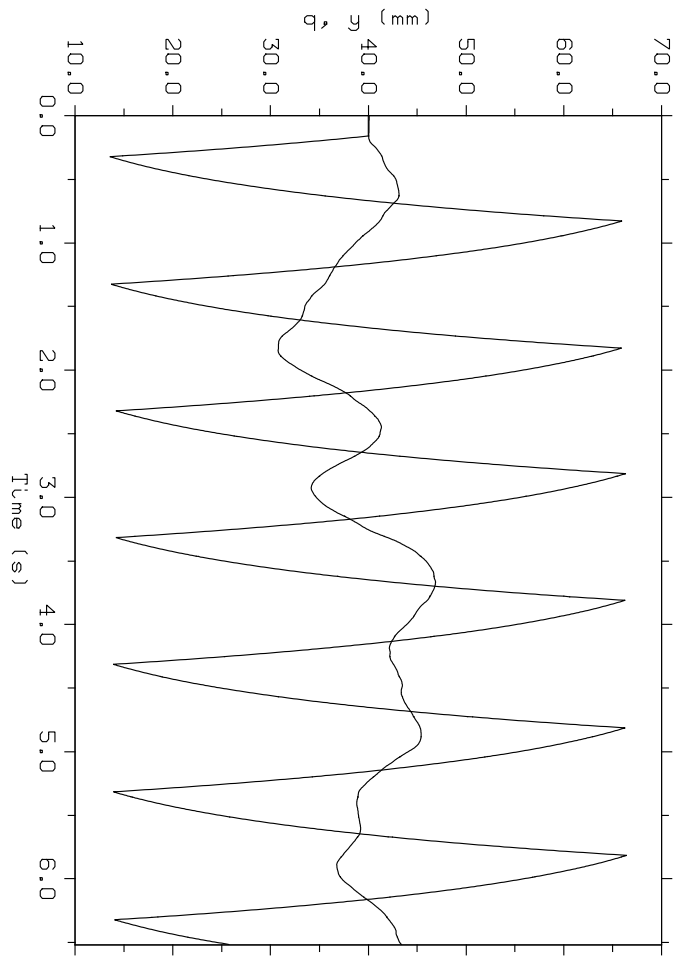}
\caption{Saw-line represents motion of the un-perturbed bang control data $q$. The wavy line represents the perturbed COG-coordinate $y$.}
\label{FIG3}
\end{figure}

\begin{figure}
\includegraphics[scale=0.6]{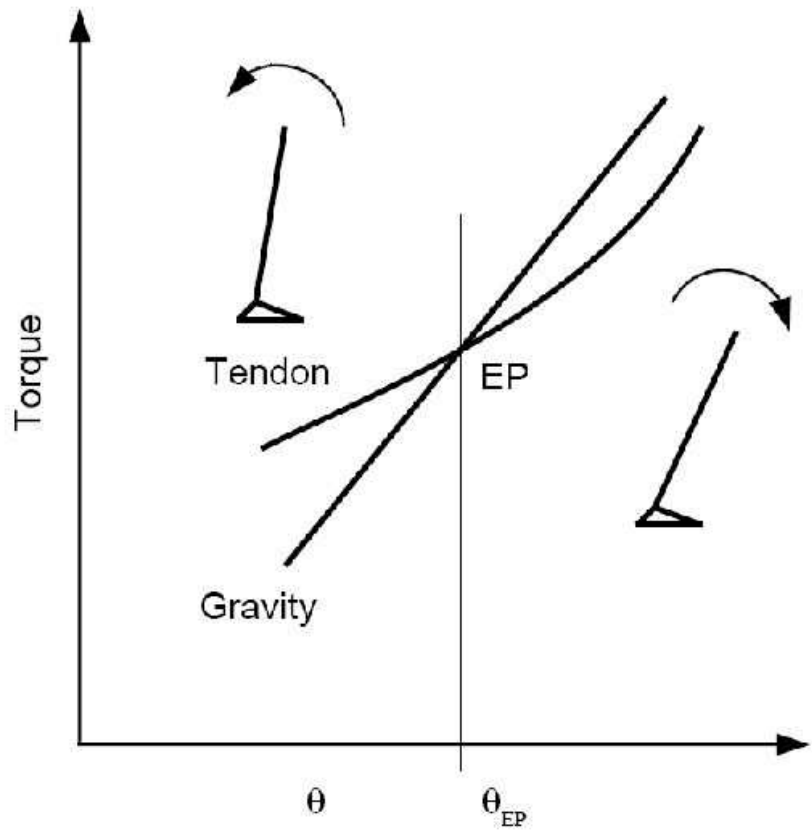}
\caption{Tendon compliance leads to instability.}
\label{FIG4}
\end{figure}

\begin{figure}
\includegraphics[scale=0.6,angle=180]{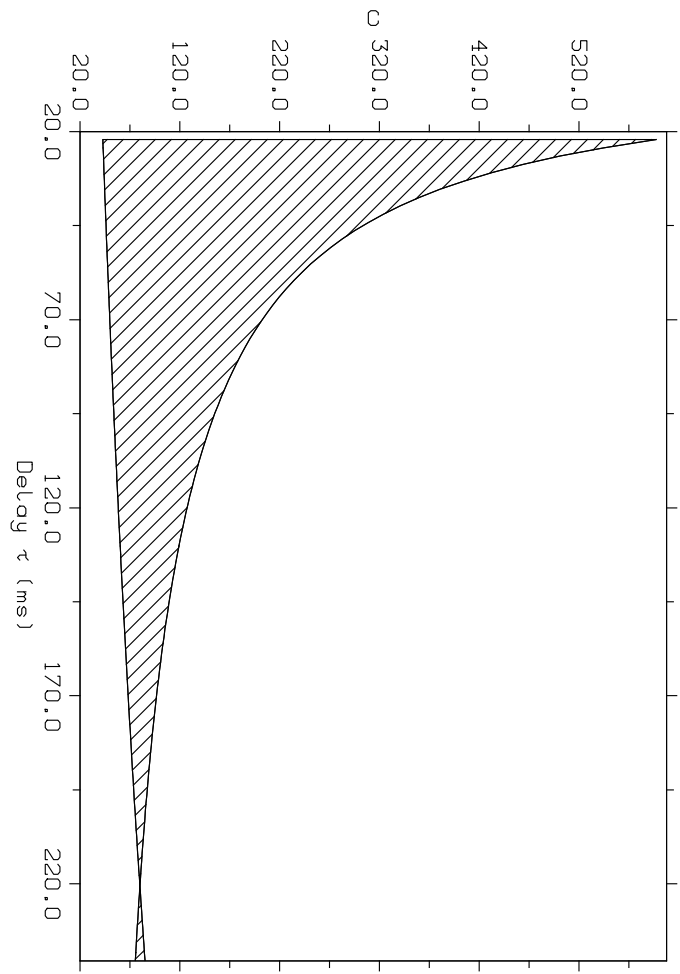}
\caption{Stability region for the bang-control.}
\label{FIG5}
\end{figure}

\begin{figure}
\includegraphics[scale=0.6]{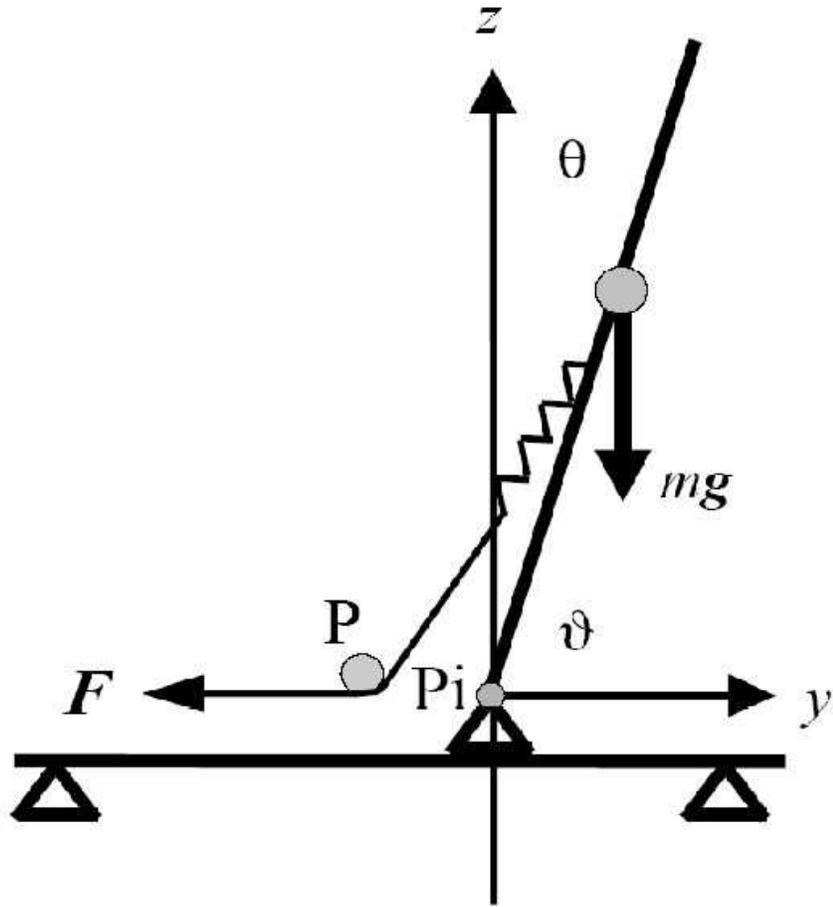}
\caption{Outline for the demonstration device.}
\label{FIG6}
\end{figure}

\end{document}